# Energy transfer and entanglement in an optically active solution of amino acids

[1]**Dipti Banerjee[1,2]**
[1]*Vidyasagar College for Women(University of Calcutta),*
*Departmentof Physics,39, Sankar Ghosh Lane, Kolkata700006,INDIA*
[2]*International Centre for theoretical Physics, Trieste, ITALY*
*deepbancu@hotmail.com*

**Abstract**: **The optical activity of a chiral medium is discussed from the view point of transfer of energy. The absorbed energy of the polarized light in the optical active medium is transferred to the mechanical rotation of the chiral molecule. They acquire the helicity dependent geometric phase due to passage of the polarized light which loses energy by having an optical rotation. The entanglement of a polarized photon and fermion is the very source of this behavior. This theoretical knowledge has been reflected in an experimental study with six essential and five non-essential amino acids.**

## 1.Introduction

Optical activity is the phenomenon by which the light matter interaction [1] is manifested through the rotation of the plane of linear polarization as light passes through chiral media. It pervades several fields of sciences and mostly an important role in the pharmaceutical industry [2]. In optical active medium the rotation of plane of polarization due to transfer of energy could be fruitfully studied in terms of angular momenta of photons [3]. Polarized light with fixed intensity has particular orbital angular momentum (OAM). Spin to orbital conversion of angular momentum is very important where variation of SAM occurring from medium's birefringence gives rise to the appearance of OAM, arising from medium's inhomogeneity. In rotationally symmetric geometries, such as "q-plates" (liquid crystal) only spin to orbital conversion [4] (STOC) process happens instead of no net development of angular momentum of light. In case of optical active solution, the rotation of polarized state on Poincare sphere indicate the change of angular momentum from one point to another. The actual nature of angular momentum transfer in optically active medium will be sorted out here.

In optically active medium with the increase of concentration of the solution, absorption of light increases. Considering six essential and five non-essential amino acids we will investigate experimentally the energy absorbed at different concentrations through optical rotation. The optical rotation of polarized photon is due to the chiral molecule in the optical active solution when there is equivalence between the optical and mechanical toque. The quantum manifestation of this mechanical torque implies that the chiral molecule will acquire a quantum phase identified as Berry phase [5]. This Berry phase had been pointed out as the natural Aharanov-Bohm phase in optically active medium acting as micro-solenoids [6].

Quantum entanglement [7,8] and quantum phases are the basic tools for quantum information processing. Quantum communication involving geometric/Berry phase plays a key role identifying Geometric quantum computation [9]. In quantum optical rotatory dispersion, the entangled photons are used to study optical activity as a function of wavelength [10]. There could be a single particle entanglement associated with macroscopic behavior. In view of recent experimental works by Kumar et.al.[11], quantum entanglement involves the biomolecule inside one living or between other neighbouring livings samples. The tied behavior of the two entities photon and fermion is an indication of entanglement. The interaction between polarized photon and chiral molecule (amino acids) has been realized through entanglement where Berry phase plays the key role. This present work would develop new approach in this area of optical activity. In the next few sections we will discuss about the physics behind optical activity in connection with angular momentum, then geometric phase and entanglement.

## 2. Optical Activity and angular momentum

The optical activity in a medium is caused by the chiral behavior of individual molecules which constraint to move electrons along helical path under the influence of the incident electric field. E. Condon [12] pointed out that, linearly polarized light (which is a superposition of two circularly polarized light) if passed through such medium, a difference in velocity of propagation of wave takes place, which further causes the difference in their refractive indices due to which the plane of polarization is rotated by an angle $\theta$ as

$$\theta = \frac{\pi(n_L - n_R)l}{\lambda} = S\, l\, c_n \quad (1)$$

where $S$ is the specific rotation and $c_n$ is the concentration of the optically active medium and $n_L$ and $n_R$ are the refractive indices for two respective chiralities (L & R).

This rotation of plane of polarization of incident light is due to transfer of energy in terms of angular momentum [13]. As polarized light is rotated in a chiral medium, the angular momentum is found to be related to the concentration of the corresponding optically active medium. Classically, observed rotation in a chiral medium depends on the number of molecules which are interacted by the polarized light, hence increase in concentration results more number of rotations. The optical rotation in the optical active medium of particular concentration could be represented in terms of difference of angular momentum for two chiralities

$$\theta = \frac{f c_n V^2}{r}\left(\frac{1}{\mathcal{L}_L} - \frac{1}{\mathcal{L}_R}\right) \quad (2)$$

where we assumed that the solution of fixed concentration is kept in a glass tube of volume $V(\pi r^2 l)$ of length $l$, radius r and $\mathcal{L}$'s are the angular momentum for left and right chiral molecules respectively.

An anisotropic medium exhibiting circular anisotropy is analogous to a rotation operator. The optical active medium could be represented by Jones matrix $R_c$ whose action with plane polarized light [14] is responsible for breaking into LCP and RCP to propagate with respective phases $\varphi_L$ and $\varphi_R$

$$R_c = \begin{bmatrix} e^{i\varphi_L} & 0 \\ 0 & e^{i\varphi_R} \end{bmatrix} \quad (3)$$

By introducing the differential phase shift $\varphi = \varphi_L - \varphi_R = \Delta n_c k_o e$, and $\theta = \varphi/2$, the matrix $R_c$ changes to R in course of change of basis vector of linear states [X,Y] to circular states [L,R]. The basis between linear states [X,Y] and the basis of the circular states [L,R] indicate the change of emerging state of polarization V' from V as V'=RV, where $\psi = (\varphi_L + \varphi_R)/2$

$$R = e^{i\psi}\begin{bmatrix} \cos\theta & -\sin\theta \\ \sin\theta & \cos\theta \end{bmatrix} \quad (4)$$

It is thus seen that ignoring the phase factor $e^{i\psi}$, the optical active medium behaves as rotation matrix (eq.4) of angle $\theta$ (eq.2). It could be mentioned here [15] that this rotator only rotates the plane of polarization of linearly polarized light by securing a geometric phase in turn where no phase for circularly polarized light. With the variation of concentration of the optically active solution, this optical rotation of plane of polarization changes indicating that the rotation matrix

$$R = \begin{pmatrix} \cos\theta & -\sin\theta \\ \sin\theta & \cos\theta \end{pmatrix} \quad (5)$$

is the source of energy absorption for linearly polarized light The Jones matrix of analyzer becomes

$$A_\alpha = \begin{bmatrix} \cos^2\alpha & \sin\alpha\cos\alpha \\ \sin\alpha\cos\alpha & \sin^2\alpha \end{bmatrix} \quad (6)$$

If the polarized light state, $X = \begin{pmatrix} 1 \\ 0 \end{pmatrix}$ from polarizer is made to pass consecutively through optical active medium (R) and analyzer $A_\alpha$, the corresponding emerging ray X' from the device will be J' = $A_\alpha$ R X. The emergent ray becomes using eqn. (5) & (6) as

$$X' = \cos(\theta - \alpha)\begin{bmatrix} \cos\alpha \\ \sin\alpha \end{bmatrix} \quad (7)$$

This angle $(\theta - \alpha) = \vartheta$ of rotation of the plane of polarization could be measured by the digital polarimeter with respect to several concentrations of the optical active medium. Photon is a two-state system. To describe polarization state in the orthogonal basis formed by the linear polarization states in the horizontal and vertical direction we may write

$|L> = \frac{1}{\sqrt{2}}(|0> + i|1>)$ and $|R> = \frac{1}{\sqrt{2}}(|0> - i|1>)$.

Hence circularly polarized states that indicate the Spin angular momentum (SAM) of polarized light are

$$|L> = \frac{1}{\sqrt{2}}\begin{pmatrix}1\\i\end{pmatrix}, |R> = \frac{1}{\sqrt{2}}\begin{pmatrix}1\\-i\end{pmatrix} \qquad (8)$$

The horizontal and vertical component of linearly polarized light can be written in terms of LCP and RCP

$$|H> = \frac{1}{\sqrt{2}}(|L> + |R>) \text{ and } |V> = \frac{1}{\sqrt{2}}(|L> - |R>) \qquad (9)$$

A horizontal polarized light |H> which is a combination of circular states [L,R] is rendered by rotation of polarization due to action of optically active medium having circular anisotropy (eq.5)

$$|H'> = R|H> = (e^{-i\theta}|L> + e^{i\theta}|R>$$
$$= (|+> + |->) \qquad (10)$$

If $I_0$ be the initial intensity of the light source, the optical rotation of the plane of polarization results the change of the intensity. According to Malus law [16], the intensity of the emergent light from analyser will be $I_\varphi = X'^*X' = I_0 \cos^2\vartheta$. For monochromatic light, the intensity is constant over a particular concentration. Thus the absorbed intensity for a particular concentration of the optical medium will be

$I' = I_0 - I_0 \cos^2\vartheta = I_0 (1-\cos\vartheta)(1+\cos\vartheta)$ depending on the final optical rotation by an angle $\vartheta$. This absorbed energy is balanced by the rotatory power by $(\theta-\alpha) = \vartheta = \rho d$ where $\rho$ being the rotatory power of the medium. This transfer of energy from light to chiral molecule is equivalent to the geometric (Berry) phase by diminishing the intensity of light. It could be realized through the interaction of linearly polarized light with the chiral molecules. The optical active medium (L or R) has fixed chiral molecules with fixed direction of helicity. With the interaction of polarized light, their direction of helicity is shifted by an angle (equivalent to optical rotation). The torque responsible for optical rotation of plane of polarization of incident linearly polarized light is balanced by the mechanical rotation of the molecule in optical active medium. In fact, this implies that the molecules will acquire a quantum mechanical phase which is equivalent to natural Berry phase as long as the polarized light passes through the solution. This idea supports the investigation of Tan [6].

In quantum information field, the interest of OAM degree of freedom of light mainly arises from the possibility of using its higher-dimensionality for encoding a large amount of information in single photons [17]. Passage of linearly polarized light through optical active medium of particular concentration, reduces its intensity or OAM along with its shift of initial polarization due to the interaction with the chiral molecule. Transfer on energy between polarized light and chiral molecule implies the transfer of OAM to SAM of molecule indicating Orbital to Spin angular momentum transfer process (OTSC).

## 3. Geometric phase and Entanglement

Quantum theory describes a physical system in terms of state vectors in a linear space. If two systems are combined then they have correlations between observations and the state is called entangled.

The chiral molecule will acquire a helicity dependent quantum phase. Quantum mechanically, this process is as similar as acquiring the Berry phase. A quantized spinor can be written [18] in use of quantum gates and an arbitrary super-position of elementary qubits |0> and |1>

$$|\uparrow,t> = [\cos\vartheta/2|0> + \sin\vartheta/2|1> e^{-i\varphi}]e^{i/2(\varphi-\chi)} \qquad (11)$$

Over a closed path, a spinor acquire the solid angle (neglecting the overall phase) as geometric phase about the quantization axis with the variation of all the three own parameters $\vartheta, \varphi$ and $\chi$.

$$\gamma_\uparrow = \oint L_{eff} dt = i \int A(\lambda) d\lambda = i \oint <\uparrow|\nabla|\uparrow> d\lambda$$
$$= i(\oint d\chi - \cos\vartheta \oint d\Phi)$$
$$= = \pi(1-\cos\vartheta) = \pi - \frac{1}{2}(e^{i\vartheta} + e^{-i\vartheta}) \qquad (12)$$

It has been pointed out earlier, [19] that this helicity dependent topological phase visualizes the physics behind the rotation about the quantization axis in the anisotropic medium where the helicity which is depicted by an angle $\chi$ shifts to $\chi+\delta\chi$ in anisotropic space-time. The molecule having fixed chirality (left or right) acquires the respective geometric phases by
$\gamma_L = \pi(1-\cos\vartheta)$ **or** $\gamma_R = \pi(1+\cos\vartheta)$.

This lead us to rewrite eq.7 as quantum state $|\Psi>$ at a particular instant t that is connected with the primary state $|\Psi_0>$ as [17]

$$|\Psi> = \cos\vartheta |\Psi_0> = 1/2\pi (\gamma_R - \gamma_L)|\Psi_0> \qquad (13)$$

indicating optical rotation originated from the difference of geometric phases for two chiralities left and right. The qubit rotation of chiral molecules could be identified as in presence of polarized light

$$|\uparrow(\theta)> = 1/2\pi(e^{i\vartheta} + e^{-i\vartheta})|\uparrow> \qquad (14)$$
$$|\downarrow(\theta)> = -1/2\pi(e^{i\vartheta} + e^{-i\vartheta})|\downarrow>$$

The shift of spin axis of the molecules are the outcome of mechanical torque balanced by the optical torque of the traversing polarized light through the optical active medium. The geometric phase gained by the polarized light due to the optical rotation ϑ will be of Pancharatnam [20]

$< X|X' > = \cos(\theta-\alpha) = \cos\vartheta = \frac{1}{2}(e^{i\vartheta} + e^{-i\vartheta})$ (15)

It is realized that if there are n number of molecules arranged in an array and light emerging from one strike another molecule then the final emergent light will be

$|H'_n> = R(n\vartheta)|H> = (R>e^{in\vartheta} + |L> e^{-in\vartheta})$ (16)

A linearly polarized light when entered into an optically active solution with specific concentration, the molecules get energy to rotate. These molecules are quantized having fixed helicity due to their own chirality and handedness either Levo or dextro of solution. Incident polarized light act as an external force to change their helicity by an small angle. This angle is the angle of optical rotation. Every molecule interacted by the polarized light acquires the Berry's geometric phase.

If two systems are combined then they have correlations between observations and the state is called entangled. An entangled state between a pair of two-level systems is a **singlet** state when the total spin =0 and the state is of the form [21].Here we intuitively realized the physics behind the concentration dependent optical rotation.In general the incident polarized photon qubit at jth state and the chiral molecule at the ith state having joint rotation by angle ϑ are tied by the entangled singlet state ofthe following form

$|\Psi_{singlet}> = \frac{1}{\sqrt{2}}(|\uparrow(\vartheta)>_1|+>_2 - |\downarrow(\vartheta)>_1|\rightarrow>_2)$

$= \frac{1}{\sqrt{2}}(|\uparrow(\vartheta)>_1 e^{i\vartheta}|R>_2 - |\downarrow(\vartheta)>_1 e^{-i\vartheta}|L>_2)$

$= \frac{1}{2\sqrt{2}}(|\uparrow>_1 e^{2i\vartheta}|R>_2 + |\downarrow>_1 e^{-2i\vartheta}|L>_2)$ (17)

This implies the non-appearance of topological phase of the chiral molecule where in the anisotropic optical medium the circular basis state [R,L] acquire the external OAM by the phase proportional to the optical rotation and the concentration of the medium.It is to be noted from ref.[15] that the concurrence, the measure of entanglement becomes cosϑ or cos2ϑ or so on which can be visualized through this geometric phase[19]. The gradual change from [|R>,|L>] , [|+>,|->] to [|+, ϑ >_j |−, ϑ >] depending on the optical rotation through concentration indicate that the shape of wave-front or the OAM of the incident light reduces without changing the SAM because the helical structure decays though the handedness remains same.

**4. Experiment with Amino Acids**

Molecules being mirror image to other called enantiomars. They usually referred to L,(plane of polarization clockwise) and D (anticlockwise rotation of plane of polarization).Mixtures of equal amount of L & D called racemic mixture.

Usually in biological samples one is active. It is to be noted that L & D forms have different biochemical and physiological properties. Objects that are not identical with mirror image are chiral. Excepts in some bacteria, the amino acids that found in several proteins are Levo. D-type amino acids are toxic and used in medicines as antibiotics. The physicochemical properties of a protein are determined by the analogous properties of the amino acids in it. Proteins are formed from many amino acids through peptide bonds. If a solution of a fibrous protein flows through a narrow tube, the elongated molecules become oriented parallel to the direction of the flow, and the solution thus becomes **birefringent / optically active**. It splits a light ray into two components that travel at different velocities and are polarized at right angles to each other.

With the above theoretical knowledge, we have studied experimentally the optical activities of some essential amino acids at first. We have prepared from every six following essential amino acids **L-Lysine, L-Leucine, L-Valine, L-Phenylalanine, L-Histidine, L-Metheonine,** eight different concentrations starting from 0.25% to 2%. Maintaining the room temperature nearly at 18℃, we at first measured for every concentration the optical rotation and the specific rotation by digital polarimeter CDP001. At first, we have measured the intensity of the Infrared laser light source using a current detector and current meter.

Incident intensity in terms of current=0.11 mA,
Source voltage= 5V
Incident power,($I_0$ )=(0.11x5) mW/sec= 0.55 mW per second.

In addition, we have taken five following non-essential amino acids **L-Alanine, L-Arganine, L-Serine , L-Glutamine** to study the absorbed intensity and geometric phase with respect to the above eight concentrations**.**

The respective intensities absorbed ($I_0 - I_0 \cos^2 \vartheta$) and geometric phases [$\pi(1-\cos\vartheta)(1+\cos\vartheta)$] are also evaluated by finding out $\cos\vartheta$. These practical findings are illustrated [22] through graphs to have a comparative study between these amino acids for different concentrations. These graphs fig1 and fig2 are drawn for eight different concentrations from 0.25% to 2% for six essential amino acids to compare absorbed intensities and geometric phase respectively. Except **L-Histidine** all the other five essential amino acids (L-Lysine, L-Leucine, L-Valine, L-Phenylalanine, L-Metheonine,) show similar nature of graph for respective geometric phase and absorbed energy. **L-Histidine** shows maximum and **L-Valine** has minimum absorption of energy at 2% concentration. Illustration of fig-3 is more prominent to realize the nature of absorption of light intensities and geometric phase for eight concentrations of six essential amino acids. With five non-essential amino acids, **L-Arganine, L-Glutamine, L-Serine, L-Alanine and L-Proline** we have repeated the experimental study for which fig-5 and fig-6 are drawn. It is seen in both the graphs that for **L-Proline** rise of absorbed energy and geometric phase is high where others show almost no variation with concentration. There remains more scope in future to work with biological samples in connection with drugs or disorders.

The chiral medium absorbs light intensity differently for LCP and RCP. The linearly polarized incident light becomes elliptical due to unequal absorption of amplitude. Reduction of intensity and wave front indicate loss of OAM that is transferred to SAM of molecules too. In quantum regime, the optical rotation of the plane of polarization will develop a phase that varies with the physical or dynamical parameters of the medium. Rotation of plane polarization results the change of helicity. With the increase of concentration more energy is absorbed. As this intensities and wave-front are representing the Orbital angular momentum (OAM) of light, the decrease of intensities due to absorption implies the reduction of OAM of Laser light. It is OAM to SAM conversion process (OTSC), where the chiral molecules gain SAM through helicity dependent Berry Phase. In the quantum aspect absorbed intensity has its origin in SAM dependent geometric phase. This idea has been verified through our study with essential amino acids.

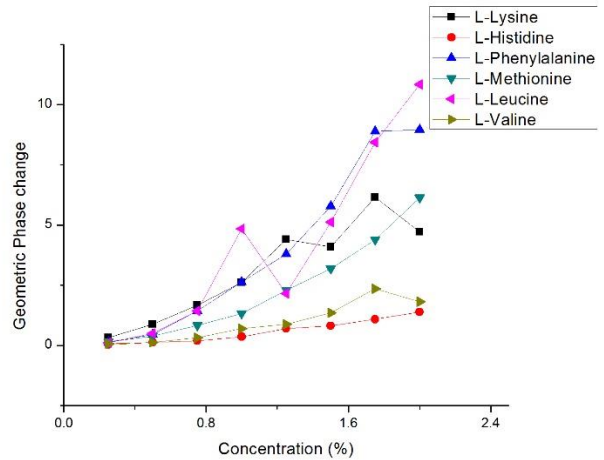

Fig. 1 : Variation of geometric phase with concentration of six essential amino acids

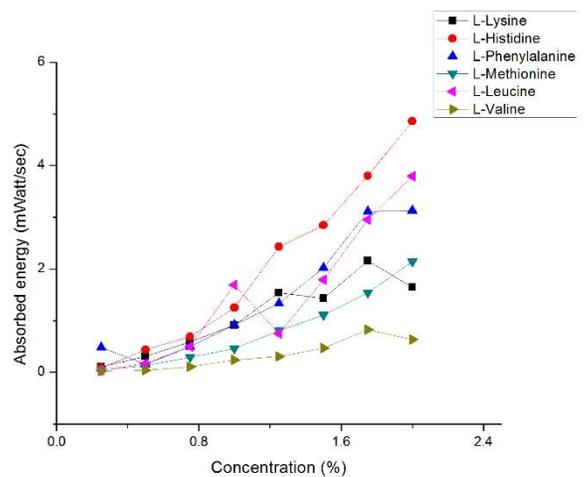

Fig. 2: Variation of absorbed energy with concentration of six essential amino acids

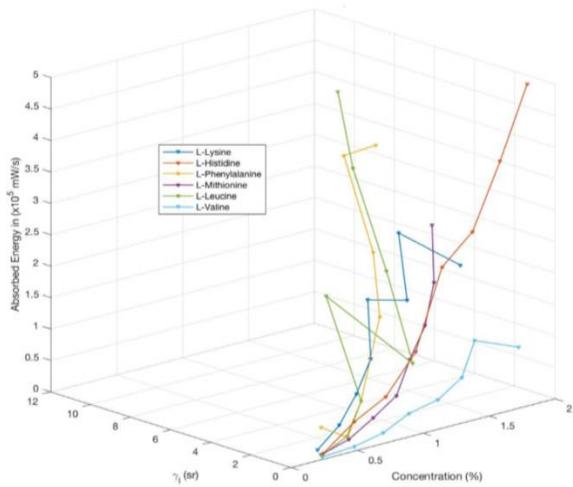

Fig. 3: Variation of absorbed energy with $\gamma_l$ and concentration of six essential amino acids

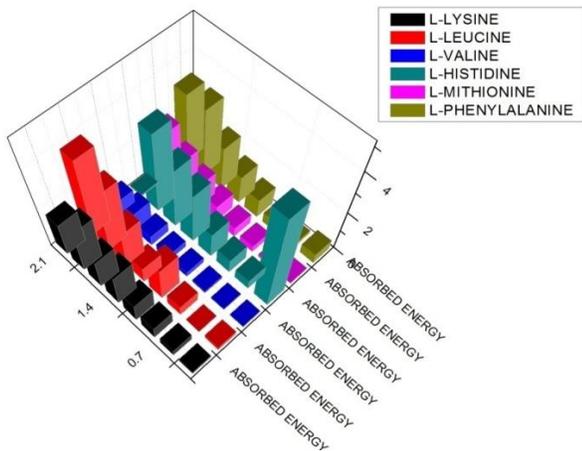

Fig. 4: Histogram of absorbed energy

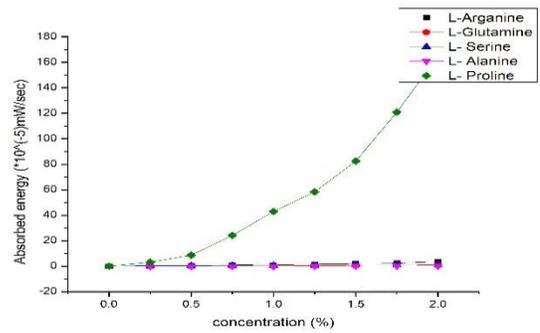

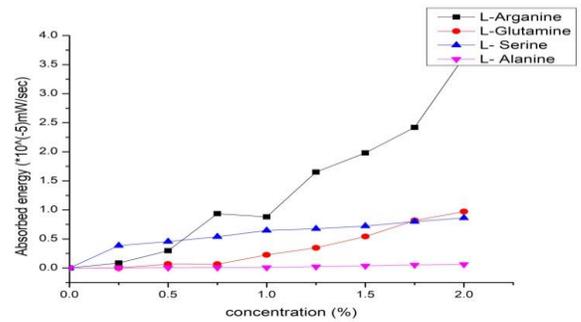

Fig.5: Variation of absorbed energy of five/four non-essential amino acids

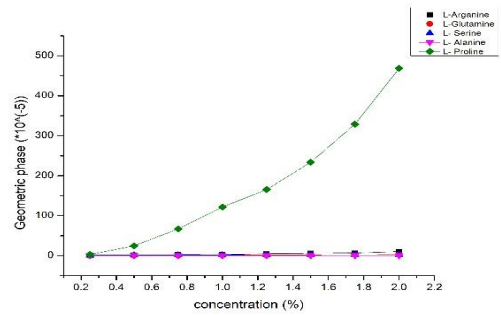

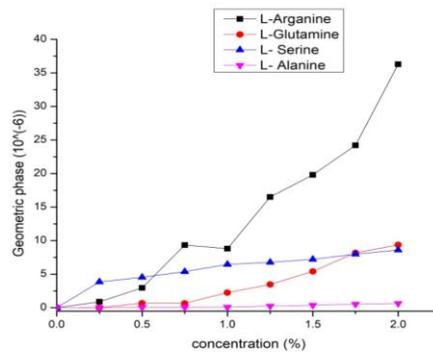

Fig.6: Variation of geometric phase with concentrations of five/four non-essential amino acids.

## 5. Conclusion

The rotation of plane of polarization of polarized light by chiral molecule is caused due to the equivalence between the optical and mechanical toque. With the increase of concentration more energy is absorbed. This implies the decrease of intensity or OAM of Laser light. It is a OTSC (OAM to SAM conversion) process, where the chiral molecules gain SAM through helicity dependent Berry Phase. In the quantum aspect absorbed intensity has its origin in SAM dependent geometric phase. This idea has been verified through our study with some essential and non-essential amino acids. Comparative graphical analysis with some amino acids are studied to understand the variation of absorbed intensity and geometric phase with respect to eight concentrations. The absorbed intensity of polarized light is the source of mechanical quantum phase due to the shift of direction of helicity. This tied behavior of the two entities photon and fermion implies the entanglement forming a singlet state. Using quantum memories for single photon entanglement with single molecule of amino acids, the quantum processing in forming further protein could be studied in future. This work could reopen new avenue involving Geometric quantum computation. With emphasis on angular momentum of optical active biological samples, quantum memory [23,24] will be of further ample interest to work. In future this work will be in support of stability of DNA, and investigation of molecular logic devices in biological systems.

**Funding Information**

This is a work under the minor research project of UGC of India having project number :ID no-WC2-157.

**Acknowledgment**.

The author acknowledge the help from all references and specially Prof. S.Shahriar of Northwestern University, Evanston, IL,USA.